\documentclass{svjour3}
\newcommand{\fant}[1]{\phantom{#1}}
\newcommand{\be}{\begin{equation}}
\newcommand{\ee}{\end{equation}}
\newcommand{\wdg}{\wedge}

\journalname{epjp}
\begin{document}

\title{Variational derivatives of gravitational actions}

\author{Ahmet BAYKAL}
\institute{Department of Physics, Faculty of Science and Letters, Ni\u gde University,  51240, Ni\u gde, Turkey
\email{abaykal@nigde.edu.tr}
}

\maketitle

\begin{abstract}
A method of calculation for the  variational derivatives for gravitational actions
in the pseudo-Riemannian case is proposed as a practical variant  of the first order formalism with constraints.
The method is then used to derive the metric field equations for  a generic $f(R)$ model.
\end{abstract}

\section{Introduction}
It is customary to start a study of  given gravitational model with a suitable action form and derive the corresponding field equations
for a model by calculating the variational derivative  chosen from a  given set of various related calculational techniques.

The so-called first order formalism is an efficient and fairly flexible calculational framework that allows one to discuss a wide class of
gravitational models ranging from modified general relativity, Riemann-Cartan type gravitational models\cite{kopczynky}, teleparallel gravitational models\cite{kopczynky2} to metric-affine
gravity \cite{hehl}. The independent gravitational field variables are basis coframe and connection forms and only the exterior
derivatives of these variables are allowed in the Lagrangian form. Within this framework, it is possible to introduce constraints for various
field variables. For example, in teleparallel gravity, the curvature is constrained to vanish whereas  in a metric theory, torsion is constrained
to vanish by introducing the appropriate Lagrange multipliers.

In the metric approach using a coordinate basis, on the other hand, the  only  gravitational variable is assumed to be the metric tensor.
The variational derivative of a given gravitational Lagrangian, which is usually composed of curvature invariants, may  then be calculated in stages from a  Lagrangian density of the form
\be
L=\mathcal{L}[g_{\alpha\beta}]*1
\ee
where $*1=\sqrt{|g_{\alpha\beta}|}\frac{1}{4!}\epsilon_{\alpha\beta\mu\nu}dx^\alpha\wdg dx^\beta\wdg dx^\mu\wdg dx^\nu$
is invariant four dimensional volume element in terms of some local coordinates $\{x^\alpha\}$.

In order to find the field equations for a gravitational Lagrangian form $L$ depending on some curvature scalars,
one calculates the variational derivative
\be
E_{\alpha\beta}
\equiv
\frac{\delta L}{\delta g_{\alpha\beta}}
\ee
in the metric formalism. The metric field equations can be found,
in the first order formalism in the manner of Palatini \cite{palatini}, with the connection constrained by introducing vanishing non-metricity
and vanishing torsion constraints relative to a coordinate basis, see for example, \cite{safko-elston}.
On the other hand, instead of introducing constraints for independent connection as a gravitational variable,
one often calculates the variational derivative
$
{\delta \mathcal{L}}/{\delta \Gamma^{\mu}_{\alpha\beta}}
$
and  then, one converts variational derivative with respect to connection coefficients (that is, the Christoffel symbols)
into the variational derivatives with respect to metric components by making use of
\be\label{delta-connection}
\delta\Gamma^{\mu}_{\alpha\beta}
=
\frac{1}{2}g^{\mu\nu}[(\delta g_{\nu\alpha})_{;\beta}+(\delta g_{\nu\beta})_{;\alpha}-(\delta g_{\alpha\beta})_{;\nu}]
\ee
and the Leibnitz property of the covariant derivatives with respect to the basis frame fields $\{\partial_\alpha \}$ denoted by $_{;\alpha}$.
In deriving (\ref{delta-connection}), one makes essential use of the classical expression
\be
\Gamma^{\mu}_{\alpha\beta}
=
\frac{1}{2}g^{\mu\nu}(g_{\nu\alpha, \beta}+g_{\nu\beta,\alpha}-g_{\alpha\beta,\nu}).
\ee

 In an analogous calculation relative to an orthonormal coframe, the metric field equations can be derived from the coframe variation in the first order formalism. The constraints  for connection can be imposed using a Lagrange  multiplier in a manner analogous to the corresponding constraints relative to a coordinate basis mentioned above. On the other hand, in a typical calculation relative to an orthonormal coframe in the pseudo-Riemannian case, the  constraints can be introduced by corresponding Lagrange multipliers and subsequently,  the Lagrange multipliers are eliminated from field equations.

The present paper introduces a method, in the pseudo-Riemannian case and using an \emph{orthonormal coframe}, which cast the constrained first order formulation in a form analogous to the well-known coordinate version highlighted above in (1)-(4).
In particular, a version of the identity in (\ref{delta-connection}) is given for the vanishing torsion constraint for a metric  compatible connection for a formulation relative to an \emph{orthonormal coframe}. This identity is then used to show that the metric field equations derived using the
identity.

For the rest of the paper, the exterior algebra of differential forms defined on pseudo-Riemannian manifolds is used. The notation used in the following
follows almost identically that of \cite{var-dereli-tucker},\cite{benn-tucker} (except the formal difference in the definition of Einstein 3-form specified below). For the convenience of the reader, the notation  can be  summarized as follows. The metric field equations for gravitational models will be derived from an action $I$ and the Lagrangian 4-form $L$ of the  form
\be
I=
\int_UL[e^a,\omega_{ab}]
\ee
defined on some open subset $U\subset M$ on a chart on a pseudo-Riemannian manifold $M$ with the metric $g=\eta_{ab}e^a\otimes e^b$.
The Lagrangian form $L$ depends on the gravitational variables $e^a,\omega_{ab}$ and their exterior derivatives.
$\{e^a\}$ is the set of $g$-orthonormal
basis 1-forms dual to the basis vector fields $X_a$. Relative to this basis the metric components are constant and $\eta_{ab}=diag(-+++)$. $i_{X_a}\equiv i_a$ denotes
the contraction operator with respect to basis vector $X_a$. The Greek letters are used for the coordinate components of tensors whereas the Latin
letters refers to the components relative to an orthonormal coframe.
The coordinate components of basis vector fields  are $e^a=e^{a}_{\fant{a}\alpha}dx^\alpha$ and $X_a=e^{\alpha}_a\partial_\alpha$ with $e^{\alpha}_{\fant{a}a}e^{b}_{\fant{b}\alpha}=\delta^{a}_{b}$, $e^{\alpha}_{\fant{a}a}e^{a}_{\fant{b}\beta}=\delta^{\alpha}_{\beta}$.
Multiple exterior products are abbreviated as $e^{a}\wdg e^{b}\wdg e^c\cdots\equiv e^{abc\cdots}$, etc. for convenience. The invariant volume element
is $*1=e^{0123}$. The Hodge dual operator  $*$ is defined by the metric and a fixed orientation and it provides an inner product
for $p$-forms. In terms of connection 1-forms $\{\omega^{a}_{\fant{a}b}\}$, the  Maurer-Cartan structure equations for the torsion 2-form $T^a$ and curvature 2-form $R^{a}_{\fant{a}b}$ read
\begin{eqnarray}
T^a
&=&
De^a
=
de^a
+
\omega^{a}_{\fant{a}b}\wdg e^b
\\
R^{a}_{\fant{a}b}
&=&
d\omega^{a}_{\fant{a}b}+\omega^{a}_{\fant{a}c}\wdg\omega^{c}_{\fant{a}b}
\end{eqnarray}
respectively. $D$ stands for the  covariant exterior derivative whereas $\nabla_a$ denotes the  covariant derivative and general definitions
can be found in for example in \cite{benn-tucker}. The Ricci 1-form and scalar  curvature can be expressed in terms of the contraction of the curvature 2-form
as $R^a\equiv i_bR^{ba}$ and $R\equiv i_a R^a$ respectively where $R^{a}_{\fant{a}b}=\frac{1}{2}R^{a}_{\fant{a}bcd}e^{cd}$.
Finally, $\delta$ stands for an infinitesimal variation in a field variable it precedes.

The organization of the paper is as follows. In  the next section, metric equations for a generic gravitational Lagrangian   are derived
from a coframe equations in the first order formalism. The pseudo-Riemannian subcase is shown to be obtained by introducing Lagrange multiplier terms
in a streamlined way. In the following  section, the  metric field equations are obtained in the same generality without the use of Lagrange multipliers
relative to \emph{an orthonormal coframe}. The proposed method of calculation for the metric equations is then applied to $f(R)$ model. The paper  concludes with brief comments on the formalism. For the sake of simplicity only the vacuum equations of the models are discussed.

\section{First order formalism in brief}
In the first order formalism, by making use of the exterior algebra of forms, it is possible to derive the orthonormal coframe
expressions for the corresponding field equations. In this more efficient method of calculations compared to coordinate calculations,
the coframe $e^a$ (metric) and the connection $\omega^{ab}$ are assumed to be independent gravitational variables where only the first order derivatives
$de^a, d\omega_{ab}$ are allowed in the Lagrangian  form $L=L[e^a, \omega_{ab}, de^a, d\omega_{ab}]$. The models with local lorentz invariance in general forbids the explicit $\omega_{ab}$ dependence of the action form and consequently $\omega_{ab}$ enters into the Lagrangian form via curvature 2-form $R_{ab}$ or its contractions.

The total variational derivative of a general  Lagrangian form $L=L[e^a, T^a, R^{ab}]$
can be written as
\be\label{total-variation-1st-order}
\delta L
=
\delta e^a\wdg \frac{\partial L}{\partial e^a}
+
\delta R^{ab}\wdg \frac{\partial L}{\partial R^{ab}}
+
\delta T^{a}\wdg \frac{\partial L}{\partial T^{a}}
\ee
where the partial derivative of the Lagrangian form with respect to a tensor-valued form can be defined in terms of usual partial derivatives and the technical details on this technical  point can be found in \cite{kopczynky}, see also \cite{hehl}.
The total variational derivative can be converted into variational derivative with respect to independent variables $\{e^a\}$ and $\{\omega_{ab}\}$
by making use of the variational identities
\be
\delta R_{ab}
=
D\delta\omega_{ab}
\ee
and
\be
\delta T^{a}
=
D\delta e^a
-
\delta\omega^{ab}\wdg \frac{1}{2}
\left(
e_a\wdg \frac{\partial L}{\partial T^{b}}
-
e_b\wdg \frac{\partial L}{\partial T^{a}}
\right)
\ee
where the coefficient of $\delta\omega_{ab}$ is antisymmetrized.
These equations yield the general variational expression
\begin{eqnarray}
\delta L
&=&
\delta e^a\wdg
\left(
\frac{\partial L}{\partial e^a}
+
D\frac{\partial L}{\partial T^{a}}
\right)
\nonumber\\
&+&
\delta \omega^{ab}\wdg
\left[
D\frac{\partial L}{\partial R^{ab}}
-
\frac{1}{2}
\left(
e_a\wdg \frac{\partial L}{\partial T^{b}}
-
e_b\wdg \frac{\partial L}{\partial T^{a}}
\right)
\right].
\end{eqnarray}
The coefficient of $\delta\omega_{ab}$ is antisymmetric in the indices $a,b$ since $R_{ab}+R_{ba}=0$ as a result of metric compatibility.
The above general expression for the variational derivative and the first order formalism  allows one to study quite diverse gravitational models within a unified variational principle, see for example, \cite{kopczynky}. The pseudo-Riemannian subcase is obtained by introducing the appropriate constraints on  the independent connection. The connection is metric compatible, which and the metric compatibility condition explicitly reads,
\be\label{zero-metricity}
D\eta_{ab}
=
-
\eta_{ac}\omega^{c}_{\fant{a}b}
-
\eta_{bc}\omega^{c}_{\fant{a}a}=0
\ee
in terms of the covariant exterior derivative relative to an orthonormal coframe. This is an algebraic constraint which can be implemented into
the variational derivative with respect to connection form simply by assuming $\delta\omega_{ab}+\delta\omega_{ba}=0$
which directly follows from (\ref{zero-metricity}).
On the other hand in order to implement zero-torsion constraint on the connection one extends the lagarangian form $L[e^a, R^{ab}]$ with a
Lagrange multiplier  term $L_C$ to have
\be
L_e[e^a, R^{ab}, T_a, \lambda^a]
=
L[e^a, R^{ab}]+L_C[T_a, \lambda^a]
\ee
where
\be
L_C
=
\lambda^a\wdg T_a
\ee
imposes the zero-torsion constraint on the connection 1-form $\omega_{ab}.$ The zero torsion constraint is
a dynamical constraint since it involves $de^a$ and $\omega_{ab}$ in the particular tensorial form
\be
T^a
=
de^a+\omega^{a}_{\fant{a}b}\wdg e^b
=
0.
\ee
The coframe (metric) equations are then take the form
\be
\frac{\delta L}{\delta e^a}
\equiv
*E^{a}
=
\frac{\partial L}{\partial e^a}
+
D\lambda^a
\ee
whereas the connection equations become
\be\label{connection-eqns}
\Pi^{ab}
=
\frac{1}{2}(e^a\wdg \lambda^b-e^b\wdg \lambda^a).
\ee
The tensor- valued auxiliary  3-form $\Pi^{ab}$, satisfying $\Pi^{ab}+\Pi^{ba}=0$  defined to be
\be
\Pi^{ab}
\equiv
D\frac{\partial L_e}{\partial R^{ab}}
=
D\frac{\partial L}{\partial R^{ab}}
\ee
since, in this particular case one has
\be
\frac{\partial L_C}{\partial R^{ab}}\equiv0.
\ee
The connection equations (\ref{connection-eqns}) can be regarded as a set of algebraic equations for the Lagrange multiplier form and they can be solved for the Lagrange multiplier $\lambda_a$ uniquely subject to the remaining field equations for the extended Lagrangian, namely,
\be
\frac{\partial L_e}{\partial \lambda_a}
=
\frac{\partial L_C}{\partial \lambda_a}
=
T^a=0.
\ee

The auxiliary tensor-valued 3-form
$\Pi^{ab}$ can have  at most 24 independent components and this number is also equal to the number of independent  components
of the Lagrange multiplier, which is a vector valued 2-form $\lambda^a=\frac{1}{2}\lambda^{a}_{\fant{a}bc}e^{bc}$. Consequently $\lambda^a$
and $\Pi^{ab}$ can be considered to be equivalent \cite{hehl}.
Explicitly, by calculating two successive contractions of the (\ref{connection-eqns}),
one can find $\lambda_a$ in terms of the other dynamical variables as
\be\label{inverted-connection-eqns}
\lambda^a
=
2i_b \Pi^{ba}
-
\frac{1}{2}e^a \wdg i_bi_c \Pi^{bc}.
\ee
Consequently, using (\ref{inverted-connection-eqns}), the variable $\lambda^a$ can be eliminated from the coframe
equations in favor of the other field variables.
In doing so, one  obtains the general form of the metric field equations
\be\label{coframe-eqns1}
*E^a
=
\frac{\partial L}{\partial e^a}
+
2
Di_b \Pi^{ba}
+
\frac{1}{2}e^a \wdg Di_bi_c \Pi^{bc}
\ee
where $E^a=E^{a}_{\fant{a}b}e^b$ and the components relative to orthonormal coframe $E_{ab}$ is related to the corresponding coordinate expression
in (2) by $E_{\alpha\beta}=e^{a}_{\fant{a}\alpha}e^{b}_{\fant{a}\beta}E_{ab}$.

Recapitulating the method of obtaining the metric equations in  the pseudo-Riemannian subcase by using the first order formalism relative to an orthonormal coframe, the coframe(i.e., the metric) equations are obtained by eliminating the Lagrange multiplier by solving the connection equations and
express the Lagrange multiplier 2-form $\lambda^a$ in terms of the remaining gravitational variables. $\lambda^a$ is calculated by taking into account
the constraint equation $T^a=0$ so that the constrained Lagrangian form $L_e$ in the constrained first order formalism corresponds  to the metric  field equations of the original Lagrangian with the metric as the only gravitational variable.

\section{Implementation of constraints without Lagrange multipliers}

The formulae (4) for the variational derivative of the Christoffel symbols can be derived from the expression (5) and
(4) relates  the variational derivative $\delta\Gamma^{\mu}_{\alpha\beta}$ to the variational derivative of the metric components relative to a local
coordinate basis. In order to dispense with the Lagrange multipliers one has to find the analogue of these relations
relative to an orthonormal coframe. This can be done as follows.

It is convenient to start with the Maurer-Cartan structure equations
\be\label{MC1-zero-T}
T^a=de^a+\omega^{a}_{\fant{a}b}\wdg e^b=0
\ee
The connection can be written in terms of basis coframe 1-forms $e^a$ and $de^a$ by solving the Maurer-Cartan structure equations with $T^a=0$.
More explicitly, by calculating the two successive contraction of the equations (\ref{MC1-zero-T})
and taking into account the metric compatibility $\omega_{ab}+\omega_{ba}=0$, it is possible to invert (\ref{MC1-zero-T}) to express $\omega_{ab}$ in the following convenient form
\be\label{on-conn-def}
\omega^{a}_{\fant{a}b}
=
\frac{1}{2}i^ai_b(de^c\wdg e_c)-i^ade_b+i_bde^a.
\ee
(\ref{on-conn-def}) is  an expression for the Levi-Civita connection corresponding to the expression (5) for the Christoffel symbols relative to a coordinate frame.

The variational derivative of (\ref{MC1-zero-T}) with $T^a=0$, yields
\be\label{MC1-zero-T2}
\delta de^a+\delta\omega^{a}_{\fant{a}b}\wdg e^b+\omega^{a}_{\fant{a}b}\wdg \delta e^b=0.
\ee
where the property that $\delta$ commutes with exterior derivative $d$ (but not with the covariant exterior derivative $D$) has been used. Rearranging the terms this can be written in the compact and convenient form
\be\label{MC1-zero-T-varied}
D\delta e^a+\delta \omega^{a}_{\fant{a}b}\wdg e^b=0.
\ee
(\ref{MC1-zero-T-varied}), in fact, relates the variational change in $\omega_{ab}$ to  $\delta e^a$. In technically the same way as
 (\ref{on-conn-def}) is derived from (\ref{MC1-zero-T}),   (\ref{MC1-zero-T-varied}) can be inverted to express $\delta\omega_{ab}$ in terms  of
 the covariant exterior derivative of the variational derivative $\delta e^a$. In effect, by noting that $\delta\omega_{ab}+\delta\omega_{ba}=0$, one can find
 \be\label{key}
 \delta \omega^{a}_{\fant{a}b}
=
\frac{1}{2}i^ai_b(D\delta e^c\wdg e_c)-i^aD\delta e_b+i_bD\delta e^a.
 \ee
This key relation  can conveniently be obtained from (\ref{on-conn-def}) by the following formal replacements
$de^a\mapsto D\delta e^a$ and $\omega_{ab}\mapsto\delta \omega_{ab}$ because  such replacements map equations (\ref{MC1-zero-T}) to equations (\ref{MC1-zero-T-varied}). (A derivation is provided in Appendix below). The relation (\ref{key}) then can be used to implement a metric variational derivative \emph{relative to an orthonormal coframe} without the need to introduce the corresponding Lagrange multiplier.
This key relation effectively eliminates the variable $\omega_{ab}$ in the metric case.
In this regard, (\ref{key}) is an expression analogous to  (\ref{delta-connection})
defined relative to a coordinate basis. In addition, (\ref{key}) obviously
allows one to adopt the basis coframe forms as the only dynamical gravitational variable.

Returning now to the total variational the expression (\ref{total-variation-1st-order}), for a general $L=L[e^a, R_{ab}]$, it can be rewritten as
\be
\delta L
=
\delta e^a\wdg \frac{\partial L}{\partial e^a}
+
\delta \omega^{ab}\wdg D\frac{\partial L}{\partial R^{ab}}.
\ee
Now, for convenience, by making using of $\Pi^{ab}$ for second the partial derivative on the right hand side  and also using
(\ref{key}) in the total variational formulae, one finds
\be
\delta L
=
\delta e^a\wdg \frac{\partial L}{\partial e^a}
+
\left[\frac{1}{2}i_ai_b(D\delta e^c\wdg e_c)-2i_aD\delta e_b\right]\wdg \Pi^{ab}
\ee
where the antisymmetry property, $\Pi^{ab}+\Pi^{ba}=0$, has been used in the second term in the square bracket.

Next, recall the following useful identity for the contraction operator
\be\label{contraction-id}
i^a\eta\wdg \sigma+(-1)^{p}\eta\wdg i^a\sigma=0
\ee
for an arbitrary  $p$-form $\eta$ and $q$-form $\sigma$ with $p+q\geq 5$. (\ref{contraction-id}) then can be used
to obtain the following relations
\begin{eqnarray}
i_aD\delta e_b\wdg \Pi^{ab}
&=&
-
\delta e_b\wdg D i_a\Pi^{ab}
+
d(e_b\wdg i_a\Pi^{ab}),
\\
i_ai_b(D\delta e^c\wdg e_c)
\wdg \Pi^{ab}
&=&
\delta e^c\wdg e_c\wdg Di_ai_b \Pi^{ab}
+
d(\delta e^c\wdg e_c\wdg i_ai_b \Pi^{ab}).
\end{eqnarray}
Using these relations, the total variational expression can in turn be rewritten  in the form
\be
\delta L
=
\delta e^a\wdg \frac{\partial L}{\partial e^a}
+
\delta e_b\wdg 2 D i_a\Pi^{ab}
+
\frac{1}{2}
\delta e^c\wdg e_c\wdg Di_ai_b \Pi^{ab}
\ee
up to irrelevant total derivative term.
This yields the total variational derivative of a given gravitational Lagrangian with respect to basis coframe  1-forms
\be\label{same}
*E^a
=
\frac{\partial L}{\partial e^a}
+
2 D i_b\Pi^{ba}
+
\frac{1}{2}
e^a\wdg Di_bi_c \Pi^{bc}
\ee
The expression on the right hand side in (\ref{same}) is the same as (\ref{coframe-eqns1}) derived previously where the dynamical constraint is introduced by using Lagrange multiplier and $D\lambda^a$ term is recovered by noting that $T^a=De^a=0$ and the linearity property of the covariant exterior derivative $D$.
It is worth to emphasize that it is obtained from the connection variational term by expressing $\delta\omega_{ab}$ in terms of $\delta e^a$.

As a result, for a general gravitational Lagrangian, the use of the expression (\ref{key}) and the subsequent well-known identities
for the contraction operator, exterior derivative and the covariant exterior derivative, one can calculate metric field equations
derived from the orthonormal coframe equations without introducing a constraint on the connection 1-form via a Lagrange multiplier term.
 The concise and practical method  of calculating metric equations makes the calculations relative to an  orthonormal coframe even more transparent
yet it also mimics the corresponding calculations relative to a coordinate basis referred in the introduction.
The advantages are in fact more evident in actual calculations for a given
gravitational Lagrangian form of any complexity. In support of this claim,  an illustrative example is presented
in some detail in the next section.

 \section{An application: A derivation of metric $f(R)$ equations}
In order to provide a concrete example regarding to the use of the general scheme of variational calculus provided above in some generality,
let us take the modified gravitational Lagrangian based on so-called $f(R)$ models \cite{ST}.
Explicitly, let us consider the  popular generalization of Einstein-Hilbert action  of the form
\be
L
=
f(R)*1
\ee
where the function $f$ is assumed to be a differentiable and arbitrary algebraic function of the scalar curvature $R$.
The total variational derivative of  the Lagrangian form then is of the form
\be
\delta L
=
\delta f(R)*1
+
f(R)\delta *1.
\ee
For the variational derivative in the first term one uses
$\delta f(R)=\frac{df}{dR}\delta R\equiv f' \delta R$
and consequently the total variational derivative can be rewritten in the form
\be
\delta L
=
\delta \omega_{ab}\wdg D(f'*e^{ab})
+
\delta e^a\wdg
[
f'R_{bc}\wdg*e^{bc}_{\fant{bc}a}
+
(f-f'R)*e_a]
\ee
Here $D$ stands for covariant exterior derivative with respect to the Levi-Civita connection $\omega_{ab}$
and  by taking into account $D*e^{ab}=0$ which follows   from the  assumption $T^a=0$ one arrives at
\be
\delta L
=
\delta \omega_{ab}\wdg df'\wdg *e^{ab}
+
\delta e^a\wdg
[
f'R_{bc}\wdg*e^{bc}_{\fant{bc}a}
+
(f-f'R)*e_a]
\ee
and thus note that $\Pi^{ab}=df'\wdg *e^{ab}$ \emph{in the Riemannian case} for $f(R)$ model.
One uses the key relation (\ref{key}) to simplify the first term on the right hand side as
\begin{eqnarray}
\delta \omega_{ab}\wdg df'\wdg *e^{ab}
&=&
\left[
\frac{1}{2}i_ai_b(D\delta e^c\wdg e_c)-i_aD\delta e_b+i_bD\delta e_a
\right]
\wdg df'\wdg *e^{ab}
\nonumber\\
&=&
-2i_aD\delta e_b\wdg df'\wdg *e^{ab}
\end{eqnarray}
because one has $i_ai_b(df'\wdg *e^{ab})=0$ identically.
Consequently, the total variational expression with respect to the basis coframe forms takes the form
\be
\delta L
=
\delta e^a
\wdg[
2D *(df'\wdg e^{a})
-
2f'*G_{a}
+
(f-f'R)*e_a]
\ee
where the definition $*G_{a}\equiv -\frac{1}{2}R_{bc}\wdg*e^{bc}_{\fant{bc}a}$ has been introduced. $G_a=G_{ab}e^b$ and $G_{ab}$ are components of the
Einstein tensor relative to  orthonormal coframe.
Consequently, the vacuum field equations $*E^a=0$ for $f(R)$ model  can be rewritten in the form
\be\label{metric-f(r)-on-version}
-
*\left(f'R^{a}-\frac{1}{2}fe^a\right)
+
D *(df'\wdg e^{a})
=0.
\ee
The second term can also be put into the well-known form by using the operator identity  $Di_a+i_aD=\nabla_a$ acting on a $p$-form \cite{benn-tucker} and by noting that
$D *(df'\wdg e^{a})=D i^a*df'$. Consequently, one ends up with
\be
D *(df'\wdg e_{a})
=
\nabla_a *df'
-
i_a(d*df').
\ee
Finally, by  noting that covariant derivative of a Riemannian connection commutes with the Hodge dual operator $*$ \cite{benn-tucker},
and that $\Delta f'\equiv *d*df'$
\be
D *(df'\wdg e_{a})
=
*[\nabla_a df'
-
(\Delta f')e_a].
\ee
Hence, the  explicit form of the fourth order equations $*E_{a}=0$ in (\ref{metric-f(r)-on-version}) is in accordance with the corresponding metric equation
\be
E_{\alpha\beta}
=
f'R_{\alpha\beta}-\frac{1}{2}g_{\alpha\beta}f-\nabla_{\alpha}\nabla_\beta f'+g_{\alpha\beta}\nabla^\mu\nabla_\mu f'=0
\ee
which can be derived relative to a coordinate basis using the formulae  (1)-(4) above.
In contrast, the derivation of the form (\ref{metric-f(r)-on-version}) is more effective
and straightforward. As advertised, the use of the algebra of exterior forms, without  introducing a constraint into the original Lagrangian  form,
together with the use of the variational identity (\ref{key}) provide a more direct means of calculation for the metric equations
in the pseudo-Riemannian case. This method of calculation was previously implied and was in fact made use of in  the limited context of Brans-Dicke theory \cite{baykal-delice-bd}.

\section{Concluding comments}

First of all, the method of calculation of the metric equations provided above  enhances the flexibility of the general first order formalism and reflects
the efficiency of the use of an orthonormal coframe instead of a coordinate coframe as well.
In that, the use of exterior algebra of tensor-valued forms and the orthonormal basis coframe 1-forms
as the basic gravitational variable  provides a considerable calculational advantage, where one has to deal with less number of indices.
Moreover, the calculations towards the elimination of the Lagrange multiplier are avoided altogether.
In contrast, a relation among (initially) independent variables that can be put into  a constraint  equation, then the constraint can be implemented into the calculation of the variational derivative by a corresponding Lagrange multiplier. Alternatively, as discussed above
it is sufficient to use the variational derivatives of the constraint equation relating the independent variables. Such an equation in principle can
simply be obtained  from the constraint itself  and subsequently it is possible to eliminate a dependent variable from the total variational derivative in this way. The idea of implementing constraints in this work is quite general and can be applied to other constrained systems as well.
The  metric field equations for any modified Lagrangian studied, for example,  in \cite{baykal-delice} can alternatively be found in this way.

As a final remark, note that
although a specific form of constraint is used to derive the field equations in two equivalent but different ways above, it
is also possible to implement the constraints  different then considered in the above example.
For instance, a constraint  enforcing the independent connection 1-form to particular field value can also be implemented into the variational procedure as well  \cite{var-dereli-tucker}. More explicitly, such a constraint can be implemented by extending the original Lagrangian form by the constraint
\be
L_C
=
\lambda_a\wdg (T^a-\mathcal{F}^a)
\ee
which  imposes the constraint $T^a=\mathcal{F}^a$ where $\mathcal{F}^a$ is given vector-valued 2-form which is a functional of some other independent field variables. In this case (\ref{MC1-zero-T-varied}) becomes
\be
D\delta e^a+\delta \omega^{a}_{\fant{a}b}\wdg e^b=\delta \mathcal{F}^a
\ee
and the inversion of this equation as well as the ensuing  equations require $\mathcal{F}^a$ to be specified in terms of the field variables and
the subsequent equations are to be changed accordingly.

\section*{Appendix}
It is possible to  derive (\ref{key})  in the same way as the equations (\ref{on-conn-def}) derived from the Maurer-Cartan structure equations. Explicitly, (\ref{on-conn-def}) can be derived from (\ref{MC1-zero-T}) as follows.

A contraction of (\ref{MC1-zero-T}) with $i_c$ yields
\be\label{app-cont1}
\omega_{ac}
=
i_cde_a+(i_c\omega_{ab})e^b
\ee
whereas a second successive contraction with $i_d$ yields
\be\label{app-cont2}
i_di_c de_a
=
-i_c\omega_{ad}+i_d\omega_{ac}.
\ee
Now by taking into account the metric compatibility condition, $\omega_{ab}+\omega_{ba}=0$
and by using (\ref{app-cont2}) in (\ref{app-cont1})  one arrives at
\be\label{app-inter-conn}
2\omega_{ab}
=
(i_{a}i_bde_c)e^c-i_ade_b+i_bde_a.
\ee
Finally, by using the double contraction identity
\be
i_{a}i_b(de_c\wdg e^c)
=
(i_{a}i_bde_c)e^c
+
i_ade_b
-
i_bde_a
\ee
(\ref{app-inter-conn}) yields the convenient form (\ref{on-conn-def}).

\end{document}